\documentclass[preprint,aps,showpacs]{revtex4}
\usepackage{epsfig}
\usepackage{graphics}
\usepackage{graphicx}

\newcommand{\Be}{\begin{equation}}
\newcommand{\Ee}{\end{equation}}
\newcommand{\Bea}{\begin{eqnarray}}
\newcommand{\Eea}{\end{eqnarray}}

\begin{document}

\title{The Effect of Pion Exchange \\  in a Relativistic Quark Model of Baryons}
\author{T.\ Goldman}\email{ tgoldman@lanl.gov}
\affiliation{Theoretical Division, MS-B283, 
Los Alamos National Laboratory, Los Alamos, NM 87545}
\author{Richard R.\ Silbar}\email
{ silbar@lanl.gov} 
\affiliation{Theoretical Division, MS-B283, 
Los Alamos National Laboratory, Los Alamos, NM 87545}

\begin{flushright}
%\vspace{-1.5in}
%{DRAFT -- PRC}\\
%January 7, 2008\\
{LA-UR-08-0168}\\
%\vspace{-0.1in}
{nucl-th/yymmnnn}\\
%\vspace*{0.5in}
\end{flushright}

\begin{abstract}

We examine the effect of adding pion exchange between quarks 
and pion self-energy corrections to the 
Los Alamos Relativistic Quark  model with a short-distance 
cutoff of the Bethe form. The contributions to the nucleon and the 
$\Delta$-baryon are small.
We conclude that the model is stable under 
this change in the sense that significant changes to the model 
parameters are not required. 

\end{abstract}

%\vspace{0.5in}

\maketitle

\section{Introduction}

The Los Alamos Relativistic Quark model (LARQ) has been applied to the 
ground states of octet and decuplet baryons~\cite{reldib}, and to small 
nuclei ($^{3}$He~\cite{relA} and $^{4}$He~\cite{GMSS}) with good numerical 
results. 
%%%%%%
The motivation is the conjecture that, just as distorted atomic electron wave functions 
can replace bond energies, electron affinities, and electronegativity, etc., and so provide 
chemistry with a quantum mechanical description of the binding of atoms into molecules, 
so too, a corresponding distortion of quark wave functions can provide an accurate 
description of the binding of nucleons into nuclei, and replace previous descriptions 
in terms of internucleon potentials or exchanges of mesons. The approach also 
implicitly provides a description of ``off-shell" nucleons, or equivalently, of the structure 
of a nucleon in a nuclear medium.
%%%%%%

For isolated baryons, the model has determined current quark wave 
functions by solving the Dirac equation with a linear scalar confining 
potential, although in principle, any other (relativistic) method for the 
calculation could be employed.  In addition, the spatial currents of single 
gluon exchange are evaluated with these wave functions to calculate the 
color-magnetic spin-spin (CMSS) interaction to the total state energy in 
its rest frame, which are essential to distinguishing the nucleon from the 
$\Delta$-baryon. A Gaussian spatial propagator with a range of order 1 fm 
is used for these gluon exchanges. 

For nuclei, a geometrically complex scalar mean field potential is introduced, 
consisting of an appropriate array of the scalar potentials truncated at 
the mid-planes between the array points. 
Variational wave functions are then employed, which consist of 
amplitude-weighted sums of the quark wave function solutions for the 
isolated potential, centered on each of the different array locations. 
The quark amplitude is assumed to be dominated 
by the contribution from a single array location (potential well) with the 
relative amplitude of the quark wave functions from the other wells taken 
as a variational parameter. 
Three quark wave functions are color-spin-isopin 
correlated to nucleon quantum numbers for each (dominant) well. 
The distributed coherent quark amplitudes are interpreted as a delocalization 
of the quarks from each nucleon. 
A final overall antisymmetrization is implemented to satisfy the Pauli principle. 
As well as the relative (delocalization) amplitude, the scale 
for the separation of the array points is used as an additional
variational parameter to find the minimum energy configuration (in a 
body-fixed frame). 
A variational estimate of the 
single-body energy is obtained as the square root of the matrix element of 
the squared Dirac Hamiltonian for the multi-well potential with these wave 
functions and the contribution of single gluon exchange to the total energy 
of the state is evaluated as discussed above. 

The effect of delocalization is as one would expect on general quantum mechanical
grounds: the kinetic energy of the quark is reduced and so likewise the overall 
energy of the state. It is of little consequence both when the potential wells are 
very close together, and also when they are far apart. In the former case as there 
is little additional volume for the delocalization to occur, and in the latter case, as 
the amplitude suppression from tunneling through the confining barrier formed by 
the truncated potentials becomes severe. Even at an intermediate well separation, 
however, delocalization does not proceed unimpeded. This is because the mass of 
a nucleon is reduced by the CMSS interaction (while it is increased %{\it cf.} the increase 
for the $\Delta$-baryon). As this interaction 
decreases with decreasing quark density, the net effect of delocalization here is an 
increase in the energy of the bound nucleon state corresponding to a repulsive 
interaction. These two opposing trends reach an equilibrium in the nuclear 
calculation and produce a modest distortion of the quark wave function for a bound 
nucleon relative to that for a free nucleon. 
%%%%%%%%%%%%%

The binding energy found for $^{3}$He is $4 \pm 2$MeV and for $^{4}$He is 
$20 \pm 6$MeV. The (uniform) well separations at these energy variational 
minima are approximately 1.8 and 1.75 fm respectively. The relative 
amplitude parameters are $\approx 0.10$ and $\approx 0.14$, corresponding to only 
a few percent difference between the probability distribution for a quark 
in the nuclear wave function compared with that for an isolated nucleon. 

One thus has a dynamical confirmation that, despite the huge strength of the 
strong interactions, something quantitatively identifiable as nucleons are the 
quasi-particles that build the nucleus, as has been assumed since the discovery 
of the neutron. This is  analogous to the construction of molecules in terms of 
identifiable atoms despite the distortion of the atomic electron structure from 
that in isolated atoms. 

When the nuclear quark wave function for $^{3}$He is applied to calculation 
of the valence quark structure function for that nucleus, the resulting EMC 
effect~\cite{EMC} (deviation from one of the ratio of the deep inelastic lepton 
scattering response function of the nucleus to that for $A$ nucleons) is found 
to be consistent~\cite{benesh} with the analysis by Kumano~\cite{Kumano} of the 
$A$-dependence of the EMC effect. This occurs with {\bf no} free parameters 
available for adjustment to match the data. 

A {\it non-relativistic} version of this model, the Quark Delocalization, Color
Screening Model (QDCSM), employs harmonic oscillator 
confining potentials that are screened~\cite{qdcsm} instead of truncated. 
It reproduces known baryon-baryon potentials~\cite{bb} as far out as the intermediate 
range ($\sim$1.5 fm). This includes the well-known intermediate-range 
nucleon-nucleon attraction, and reproduces the experimental $^1$S$_0$ 
scattering phase shift~\cite{ps}. 
At larger distances, however, 
the interaction potential decreases at a Gaussian rate rather than exponentially, 
due to the Gaussian form of the quark wave functions. Since the scale for the 
Gaussian fall-off is $\sim 1$ fm, the model does not describe the deuteron at all 
well. An extension to this model was implemented~\cite{xtnded} by adding 
pion exchange between quarks with a short-distance cutoff. 
The rationale for the cutoff is that all meson exchanges at  
short distances are already taken into account by delocalization of the quark wave 
functions.  With a cutoff of order $0.6 - 1.0$ fm, a deuteron of approximately normal 
size is obtained with a value for the ratio of D- to S-wave components of 4\%, 
comparable to the normally reported 5\%. 
Improvements to the $^3$S$_1$ and $^3$P$_1$ phase shifts are also found.

We therefore consider here the effect of adding pion exchange between quarks 
to the {\it relativistic} LARQ model,  also with a short-distance cutoff, of the Bethe 
form~\cite{bethe47}. 
This effectively gives quarks a form factor.
For consistency with the (broken) chiral symmetry of Quantum 
ChromoDynamics (QCD), we must also include the change in the self-energy due 
to pion-loop corrections to the quark propagator. The form factor amounts to 
converting (for this purpose) the current quarks of the LARQ to constituent quarks 
with a finite structural size: the energy takes the place of the mass although the 
wave functions remain unaltered from those of the massless current quarks. 

The outline of the paper is as follows: In Sec.\ II we calculate the contribution 
of pion exchange between different quarks in the baryon wave function. In 
Sec.\ III, for consistency with chiral theory, we calculate the contribution of 
pion loops to the quark self-energy in three approximations: 1) no excitation 
of the bound-state quark to a higher mode (as in Ref.~\cite{Inoue}), and 
2) two variations with a free 
quark propagator in the intermediate state.  The first allows comparison with 
previous work. As none of these results is large, we conclude that the size of this 
contribution is not significant. Finally, in Sec.\ IV, we present 
a discussion, including the relation to chiral perturbation theory, 
and our conclusions. 

%%%%
%\pagebreak
%\begin{flushright}
%{\today}\\
%\end{flushright}

\section{Pion-exchange between different quarks}\label{sec:piexch}

The contribution to the self energy of the nucleon (or the $\Delta$ resonance) 
due to the exchange of a pion between any two of its three valence quarks, 
labelled  ``1'' and ``2'' , is~\cite{FGP}
\Be				
  B^{(1,2)}_{q q \pi}(r_{\rm cut}) = - g^2_{q q \pi} 
     \int d^3 r_1 \int d^3 r_2 \frac{\delta_{jk} e^{-m_\pi r}}{4 \pi r} \Theta(r - r_{\rm cut}) 
     (\bar{\psi}_1 i \gamma _5 \tau_j \psi_1) (\bar{\psi}_2 i \gamma _5 \tau_k \psi_2) 
      \ . \label{eq:PPS}			
\Ee
Note the lack of a ``crossed'' term corresponding to $1 \rightarrow 2$ and 
$2 \rightarrow 1$.
This is because the quarks in a baryon are distinguishable, necessarily having different
colors.  
(We suppress here, and in the following, the antisymmetric color wave function.)
We shall discuss the value of the coupling constant $g^2_{q q \pi}$ in detail below.
The pion propagator depends on $r = |{\bf r}_1 - {\bf r}_2|$, the distance
between the two quarks.
The $S$-wave quark wave function $\psi_1$ has the form
\Be
 \psi_{1}({\bf r}_1) = \left( \begin{array}{c}
           \psi_a(r_1) \\
	   -i \mbox{\boldmath$\sigma$} \cdot {\bf \hat{r}}_1 \; \psi_b(r_1) 
	   \end{array} \right) \chi_{s_1} \xi_{t_1} \  ,
  \label{eq:psicolvec}
\Ee
with spin and isospin functions $\chi$ and $\xi$.
A similar expression holds for $\psi_{2}$.

As the LARQ models the contribution of all meson-exchange effects at short 
distances, we are only
interested in finding the contribution from pion exchange at large distances.
Equation (\ref{eq:PPS}) contains a $\Theta$-function to assure that
the six-dimensional integral there is to be evaluated for 
$r \geq r_{\rm cut} \approx 1$ fm. 
For example, the most recent Nijmegen~\cite{nij} 
fits to the $NN$ scattering data cut the unambiguous pion-exchange contribution to the 
analysis at $r_{\rm cut} = $ 1.4 fm.

\subsection{Quark Wavefunctions}

The radial wave functions $\psi_{a,b}$ in Eq.\ (\ref{eq:psicolvec}) were obtained by
solving the Dirac equation for massless quarks in a linear scalar 
potential of the form~\cite{GMSS,Critch}
\Be
   V(r) = \kappa^2 (r - r_0) \; , \label{eq:potnl} 
\Ee
where the negative offset, $-\kappa^2 r_0$, effectively gives
these massless quarks a constituent mass.  
(It also affords inclusion of a rough representation of the effect of the 
short-distance color Coulomb interaction between quarks.)
For the values used in Goldman, Maltman, Stephenson and Schmidt~\cite{GMSS} 
(referred to henceforth as GMSS), $\kappa^2 = 0.9$ GeV/fm and $r_0 = 0.57$ fm, 
the bound quark has a total energy of 361.6 MeV .
The upper and lower radial wave functions, $\psi_a(r)$ and $\psi_b(r)$, 
are chosen real in this description.

The radial functions $\psi_{a,b}(r)$ were found in GMSS
by numerically solving the Dirac equation with the linear scalar potential
of Eq.\ (\ref{eq:potnl}). In this form, 
they have dimensions ${\rm fm}^{-3/2}$ and are
normalized so that
\Be
	\int d^3 r \; \psi^\dagger({\bf r}) \psi({\bf r}) = 
	4 \pi \! \int_0^\infty r^2 dr \; [\psi_a^2(r) + \psi_b^2(r)] = 1
	\label{eq:psinorm}
\Ee
Following Critchfield \cite{Critch}, we can express the  $\psi_{a,b}(r)$
in terms of dimensionless, unnormalized functions
$\phi_{a,b}(\kappa r)$, where $\kappa = 2.1357$ $ {\rm fm}^{-1}$
(the GMSS value), as
\Be
	\psi_{a,b}(r) = N \; \kappa^{3/2}\phi_{a,b}(\kappa r)  \; .
	\label{eq:normpsihat}
\Ee
Explicitly separating out the natural scale of the problem, $\kappa$,
allows us to define the normalization factor, $N$, to be dimensionless also.

The numerical solutions for $\phi_{a,b}(x)$, $x = \kappa r$,
have been well fitted (in GMSS's Appendix B) as a sum of Gaussians,
\Bea
  \phi_a(x) &=& \sum_{k_1 = 1}^{n} a_{k_1} 
	\exp(-\tilde{\mu}_{k_1} x^2 /2) \; , \label{eq:psihata} \\
  \phi_b(x) &=& x \sum_{k_2 = 1}^{n} b_{k_2}
	\exp(-\tilde{\mu}_{k_2} x^2 /2) \; , \label{eq:psihatb}
\Eea
where we have defined $\tilde{\mu}_k = \mu_k + 1$.  
The (dimensionless) coefficients $a_k$, $b_k$, and $\mu_k$ are 
tabulated in Appendix B of GMSS  for  $n = 6$, 12, and 18 Gaussian terms.
For our purposes, the table for six terms provides sufficient accuracy.
With Eqs.\ (\ref{eq:psihata}) and (\ref{eq:psihatb}) for $\phi_{a,b}$,  
we can calculate the normalization constant, finding
\Be
  N^{-2} = 4 \pi \int_0^\infty x^2 dx \;
	[\phi_a^2(x) + \phi_b^2(x)] 
	= 170.37 = (13.05)^2  \  . 	\label{eq:normzn}
\Ee

\subsection{The Interaction Energy} 

Let us now calculate the interaction energy $B^{(1,2)}_{q q \pi}(r_{\rm cut})$
between these two quarks.
With Eq.\ (\ref{eq:psicolvec}), the two wave-function factors we need are
\Bea
  \bar{\psi}_1 i \gamma_5 \tau_j \psi_1 &=& 
      2 \psi_a(r_1) \psi_b(r_1) \; \hat{r}_{1,\alpha} \ 
    (\chi^{\dagger}_{s_{1f}} \; \sigma_\alpha  \; \chi_{s_{1i}})    
    \ (\xi^{\dagger}_{t_{1f}} \tau_j \;\xi_{t_{1i}}) \ , \label{eq:psibar1gamma5psi1} \\
  \bar{\psi}_2 i \gamma_5 \tau_k \psi_2 &=& 
      2 \psi_a(r_2) \psi_b(r_2) \; \hat{r}_{2,\beta}\ 
    (\chi^{\dagger}_{s_{2f}} \; \sigma_\beta \; \chi_{s_{2i}})     
    \ (\xi^{\dagger}_{t_{2f}} \tau_k \;\xi_{t_{2i}}) \  .
  \label{eq:psibar2gamma5psi2}
\Eea
Thus, 
\Be
   B_{q q \pi}^{(1,2)} = -2 \; (\frac{g^2_{q q \pi}}{4\pi})  \;  
      (\chi^{\dagger}_{s_{1f}} \sigma_{\alpha} \; \chi_{s_{1i}}) \;
      (\chi^{\dagger}_{s_{2f}} \sigma_{\beta} \; \chi_{s_{2i}}) \;
      (\xi^{\dagger}_{t_{1f}} \mbox{\boldmath$\tau$} \; \xi_{t_{1i}}) \cdot 
      (\xi^{\dagger}_{t_{2f}} \mbox{\boldmath$\tau$} \; \xi_{t_{2i}}) \;\;
      I_{\alpha\beta} \; ,
  \label{eq:newdirectPPS}
\Ee
where
\Bea
  I_{\alpha\beta}(r_{\rm cut}) &=& 
     \int d^3 r_1 \int d^3 r_2 \;{\bf \hat{r}}_{1,\alpha}  {\bf \hat{r}}_{2,\beta} \;
        \psi_a(r_1) \psi_b(r_1)\; \frac{e^{-m_\pi r}}{r}  \Theta(r - r_{\rm cut})
        \; \psi_a(r_2) \psi_b(r_2)  \nonumber \\ 
     = N^4 \!\! &\kappa& \int d^3 x_1 \int d^3 x_2 \;
	{\bf \hat{x}}_{1,\alpha}  {\bf \hat{x}}_{2,\beta} \;
        \phi_a(x_1) \phi_b(x_1)\; \frac{e^{-\mu x}}{x}  \Theta(x - x_{\rm cut})
        \; \phi_a(x_2) \phi_b(x_2) \ .
  \label{eq:Idirectalphabetadef}
\Eea
Here $\mu = m_\pi/\kappa$, ${\bf x}_1 = \kappa {\bf r}_1$, ${\bf x}_2 = \kappa {\bf r}_2$,
$x = |{\bf x}_1 - {\bf x}_2|$, and $x_{\rm cut} = \kappa r_{\rm cut}$.

Since all vectors in Eq.\ (\ref{eq:Idirectalphabetadef}) are integrated out, the tensor $I_{\alpha\beta}$ reduces to
\Be
  I_{\alpha\beta}(r_{\rm cut}) = 
    \delta_{\alpha\beta} \;\; I(r_{\rm cut}) \label{eq:IabIdirectexch} \ ,
\Ee
where
\Be
  I(r_{\rm cut}) = \frac{N^4 \kappa}{3} 
	\int d^3 x_1 \int d^3 x_2 \;{\bf \hat{x}}_1 \cdot {\bf \hat{x}}_2 \;
	\phi_a(x_1) \phi_b(x_1)\; \frac{e^{-\mu x}}{x} \Theta(x - x_{\rm cut})
	\; \phi_a(x_2) \phi_b(x_2) \ \label{eq:IexchTrace} .
\Ee
Physically, this simplification happens because the pion-exchange interaction between two 
spin-$\frac{1}{2}$ particles in a relative $S$-wave can only involve the central scalar, 
but not the tensor, contribution.

%%%%%

\subsection{Evaluating the Integral $I(r_{\rm cut})$}

To evaluate the integral of Eq.\ (\ref{eq:IexchTrace}), we will do the ${\bf x}_2$ integration first, rewriting
\Be
  I(r_{\rm cut}) = \frac{N^4 \kappa}{3}
     \int d^3 x_1 \; \phi_a(x_1) \phi_b(x_1) \; 
     J({\bf x}_1, x_{\rm cut}) \ , \label{eq:Idirectexchdef} 
\Ee
where
\Be
  J({\bf x}_1, x_{\rm cut}) =
     \int d^3 x_2 \; \phi_a(x_2) \phi_b(x_2)\; {\bf \hat{x}}_1 \cdot {\bf \hat{x}}_2 \;
     \frac{e^{-\mu x}}{x} \Theta(x - x_{\rm cut})\ .
  \label{eq:Jdirdef}
\Ee
To proceed, let the integration variable for $J({\bf x}_1, x_{\rm cut})$ 
be ${\bf x} = {\bf x}_1 - {\bf x}_2$, with $x$ running from $x_{\rm cut}$ to $\infty$.
As Eq.\ (\ref{eq:Jdirdef}) has no vectorial dependence remaining after the integrations,
\Bea
  J({\bf x}_1,x_{\rm cut}) &=&   J(x_1,x_{\rm cut})  \nonumber \\
	= 2 \pi &&\!\!\!\!\!\!\!\!\!\!
	\int_{x_{\rm cut}}^\infty x^2 dx \int_{-1}^1 d \cos\theta \; \;
	\frac{(x_1 - {\bf \hat{x}}_1\cdot{\bf x})} {|{\bf x}_1 - {\bf x}|} \;
	\frac{e^{-\mu x}}{x} \;
	\phi_a(|{\bf x}_1 - {\bf x}|) \phi_b(|{\bf x}_1 - {\bf x}|) \label{eq:Jx1xcut} \ ,
\Eea 
with $\cos\theta = {\bf \hat{x}}_1 \cdot {\bf \hat{x}}$.
Defining 
\Be
	c_{k_1 k_2} = (\tilde{\mu}_{k_1} + \tilde{\mu}_{k_2})/2 \label{eq:ck1k2def}
\Ee
and using Eqs.\ (\ref{eq:psihata}) and (\ref{eq:psihatb}),
\Be
  \phi_a(|{\bf x}_1 - {\bf x}|) \phi_b(|{\bf x}_1 - {\bf x}|) = 
       |{\bf x}_1 - {\bf x}| \sum_{k_1, k_2} a_{k_1} b_{k_2} 
	e^{-c_{k_1 k_2} |{\bf x}_1 - {\bf x}|^2} \ \label{eq:psiuplosum} .
\Ee
The factor of $|{\bf x}_1 - {\bf x}|$ here cancels the denominator in Eq.\ (\ref{eq:Jx1xcut}).
Expanding $|{\bf x}_1 - {\bf x}|^2$ as $x_1^2 + x^2 - 2 x_1 x \cos\theta$, we can write
\Be
  J(x_1,x_{\rm cut}) = 
	\sum_{k_1, k_2} a_{k_1} b_{k_2} \;
	J_{k_1 k_2}(x_1,x_{\rm cut}) \ \label{eq:Jassum} ,
\Ee
where
\Be
  J_{k_1 k_2}(x_1,x_{\rm cut}) = 
	\int_{x_{\rm cut}}^\infty dx \; \int_{-1}^1 d \cos\theta \;
	x(x_1 - x \cos\theta) \; e^{-\mu x - c_{k_1 k_2} (x_1^2 + x^2 - 2 x_1 x \cos\theta)} \ \label{eq:Jkkdef} .
\Ee

The double integral in Eq.\ (\ref{eq:Jkkdef}) can be done analytically, yielding
\newcommand{\erfc}{{\rm erfc}}
\Be
   J_{k_1 k_2}(x_1,x_{\rm cut}) = \frac{1}{8 x_1^2} 
     [A_{k_1 k_2}(x_1,x_{\rm cut}) - B_{k_1 k_2}(x_1,x_{\rm cut})] \ \label{eq:JasAmiB} , 
\Ee
where
\Be
   A_{k_1 k_2}(x_1,x_{\rm cut}) = \frac{2 x_1}{c_{k_1 k_2}^2} \; e^{-\mu x_{\rm cut}} \;
     [e^{-c_{k_1 k_2}(x_1 + x_{\rm cut})^2} + e^{-c_{k_1 k_2}(x_1 - x_{\rm cut})^2}] \ \label{eq:Akk}  
\Ee
and
\Bea
   B_{k_1 k_2}(x_1,x_{\rm cut}) &= & 
      \left( \frac{\pi}{c_{k_1 k_2}^5} \right)^{1/2} e^{+\mu^2/4 c_{k_1 k_2}} 
      \{ (1 + \mu x_1) \; e^{-\mu x_1} \; \erfc[t_{k_1 k_2}^{(+)}] \nonumber \\
      & & \qquad\qquad\qquad\qquad\qquad
       - (1 - \mu x_1) \; e^{+\mu x_1} \; \erfc[t_{k_1 k_2}^{(-)}] \} 
     \ \label{eq:Bkk} .
\Eea
(For definitions of the error functions, see Ref. \cite{AandS}.)
In the equation for $B_{k_1 k_2}$ we have defined  %   (x_1, x_{\rm cut})
\Be
   t_{k_1 k_2}^{(\pm)}(x_1, x_{\rm cut})  = 
      \sqrt{c_{k_1 k_2}} (x_{\rm cut} \mp x_1) + \mu/(2 \sqrt{c_{k_1 k_2}}) \; . \label{eq:tkkpm}
\Ee
One might worry that the integration over $x_1$ that follows in Eq.\ (\ref{eq:Idirectexchdef}) 
would blow up because of the $e^{+\mu x_1}$ in the second term of $B_{k_1 k_2}$. 
However, this exponential growth will be damped by the Gaussians in the 
$\phi_a(x_1) \phi_b(x_1)$ factor.

We have at this point reduced the six-fold integration in Eq.\ (\ref{eq:newdirectPPS})
to 36 simple, compact integrations over $x_1$, which we can easily (and quickly) evaluate numerically.
\Be
   I(r_{\rm cut}) 
      =  \frac{\pi^2 N^4 \kappa}{3} \sum_{k_1,k_2 = 1}^6 a_{k_1} b_{k_2} \; \;
         I_{k_1 k_2}(r_{\rm cut}) \ , \label{eq:Isum}   
\Ee
where
\Be
  I_{k_1 k_2}(r_{\rm cut}) = 
	\int_0^\infty d x_1 \; \phi_a(x_1) \phi_b(x_1) \;
	[A_{k_1 k_2}(x_1,x_{\rm cut}) - B_{k_1 k_2}(x_1,x_{\rm cut})] \ \label{eq:finalIexch} .
\Ee

We have done the numerical integrations using Mathematica~\cite{MMa}, which has 
a built-in $\erfc(x)$ function.  
For $r_{\rm cut} = 1.0$ fm, for example, $\; I = -0.487$ MeV.
The dependence of $\; I(r_{\rm cut})$ on $r_{\rm cut}$ is displayed in 
Fig.\ \ref{fig:Ircut}.
Note that this function is repulsive for $r_{\rm cut} < 0.6$ fm and attractive after that.
This can be understood as due to the interplay between the cosine from
${\bf \hat{r}}_{1} \cdot {\bf \hat{r}}_{2}$ and the cutoff function $\Theta(r - r_{\rm cut})$.
When $r_{\rm cut}$ is small, the angle between ${\bf \hat{r}}_{1}$ and ${\bf \hat{r}}_{2}$
can also be small enough so the cosine is positive.
This combined with the resulting smaller $r$ means that the positive pion propagator factor
can give larger contributions to the integrand.
However, when $r_{\rm cut}$ gets sufficiently large, the two vectors will tend to contribute
when they point in opposite directions.
This makes the cosine negative, which makes the integrand negative for $r_{\rm cut} > 0.6$ fm,
even though the pion propagator factor is diminished in size.

As a check on our algebra, we have also calculated the {\it triple} integral in Eq.\ (\ref{eq:Idirectexchdef})
for a few cases of $r_{\rm cut}$ numerically.
This takes, of course, considerably more computing time and the calculation is slow to converge.
However, the numerical results are in good agreement with the results shown in Fig.\ \ref{fig:Ircut}.

\subsection{Spin-Isospin Dependence of the Interaction Energy \label{sec:XST}}

Altogether, contribution to the nucleon or $\Delta$ mass from the interaction energy of 
cut-off pion exchange between quarks 1 and 2 is
\Be
   B^{(1,2)}_{q q \pi} = - 4\;(\frac{g^2_{q q \pi}}{4\pi}) \;\; 
	 I(r_{\rm cut}) \;\;  X^{(1,2)} \label{eq:fullBhat12}  \label{eq:fullBcheck12}
\Ee
with the spin-isospin factor
\Bea
    X^{(1,2)} = 
      (\chi^{\dagger}_{s_{1f}} \mbox{\boldmath$\sigma$} \; \chi_{s_{1i}}) \cdot
      (\chi^{\dagger}_{s_{2f}} \mbox{\boldmath$\sigma$}  \; \chi_{s_{2i}}) \;\;
      (\xi^{\dagger}_{t_{1f}} \mbox{\boldmath$\tau$} \; \xi_{t_{1i}}) \cdot 
      (\xi^{\dagger}_{t_{2f}} \mbox{\boldmath$\tau$} \; \xi_{t_{2i}}) \ .
      \label{eq:X12}
\Eea

Summing up over all pairs of the three quarks, the total $\pi$-exchange 
interaction energy contribution, after taking into account the coupling of the quark spins and
isospins to total $S$ and $T$, is
\Be
   B_{q q \pi}^{\rm tot} = - 4\;(\frac{g^2_{q q \pi}}{4\pi}) \; 
	I(r_{\rm cut}) \; X_{ST} \ , \label{eq:fullBtot} 
\Ee
where
\Be
   X_{ST} = \sum_{i<j}  
     <S,M_S | \mbox{\boldmath$\sigma$}^{(i)}\cdot\mbox{\boldmath$\sigma$}^{(j)} |S,M_S>
     <T,M_T | \mbox{\boldmath$\tau$}^{(i)}\cdot\mbox{\boldmath$\tau$}^{(j)} |T,M_T> \  . \label{eq:XSTdef} 
\Ee
 
The spin-isospin sum $X_{ST}$ can be evaluated in several ways.
The most direct (most simplistic) is to write the operators
$\mbox{\boldmath$\tau$}_i\cdot\mbox{\boldmath$\tau$}_j$ and
$\mbox{\boldmath$\sigma$}_i\cdot\mbox{\boldmath$\sigma$}_j$ in terms of their raising, lowering,
and the $z$-components.
Then apply these expanded operators, in turn, for each
combination of $i < j$ on the three-quark wave functions for the nucleon and $\Delta$.  
This proceeds quickly for the $\Delta$ state of highest weight, $|\Delta^{++},+3/2>$,
and gives, after summing, the value 3.  
For the proton with spin-up, $|p,+1/2>$, however, it is a rather tedious calculation
which eventually yields the value 15.

A more group-theoretic way of arriving at these numbers is given in an Appendix.  
However, a simple, heuristic derivation of $X_{ST}$ goes as follows.
First note that, for any pair of quarks,
\Be
   <\mbox{\boldmath$\sigma$}_i\cdot\mbox{\boldmath$\sigma$}_j> = 4<{\bf s}_i \cdot {\bf s}_j>
   = 2 S(S+1) - 3 \  , \nonumber
\Ee
which is equal to $+1$ when the two quarks are in a triplet state and $-3$ when in a singlet.
Likewise, $\mbox{\boldmath$\tau$}_i\cdot\mbox{\boldmath$\tau$}_j$ is $+1$ and $-3$ in
iso-triplet and iso-singlet states.
As mentioned earlier, each $S$-wave quark pair must be symmetric in the spin-isospin space.

For the $\Delta$, therefore, to get the total spin and isospin to both be $3/2$,
each quark pair must have $S = T = 1$.
Thus 
\Be
   <\mbox{\boldmath$\sigma$}_i\cdot\mbox{\boldmath$\sigma$}_j>
   <\mbox{\boldmath$\tau$}_i\cdot\mbox{\boldmath$\tau$}_j> = (+1)(+1) = 1 \ 
\Ee
and summing over the three pairs then yields 3 for the value of $\Delta$'s $X_{ST}$.

For the nucleon, however, each quark pair can be in either an $S=T=1$ or an $S=T=0$ state, 
and it is equally likely to be one or the other.  Thus
\Be
   <\mbox{\boldmath$\sigma$}_i\cdot\mbox{\boldmath$\sigma$}_j>
   <\mbox{\boldmath$\tau$}_i\cdot\mbox{\boldmath$\tau$}_j> = 
   \frac{1}{2}(+1)(+1) + \frac{1}{2}(-3)(-3) = 5 \ .
\Ee
Now summing over the three pairs yields 15 for the value of nucleon's $X_{ST}$.

Note that, between two quarks, the spin-isospin factor is either +1 ($S = T = 1$) or
+9 ($S = T = 0$).
Thus,  for $r_{\rm cut} > $ 0.7 fm (see Fig.\ \ref{fig:Ircut}), Eq.\ (\ref{eq:fullBtot}) shows that
the net interaction energy, $B_{q q \pi}^{(1,2)}$, is positive. 
This result is to be contrasted with the case of two nucleons in an $S$-wave. 
In that case there is no color quantum number to provide for antisymmetrization.
Thus antisymmetry of the two 
fermions involved requires that $S = 0$ and $T = 1$ or $S = 1$  and $T = 0$ (deuteron), 
and hence the interaction energy from the pion exchange has the opposite sign: 
$(-3) \times (+1) = (+1) \times (-3) = -3$. 
Thus, while this pion-exchange interaction is attractive in both
the deuteron and threshold state, it is repulsive between quarks in both the nucleon 
and the $\Delta$-baryon, as one expects for a self-energy contribution to a stable particle. 

\subsection{Evaluation of the Full Exchange Energy $B^{\rm tot}_{qq\pi}(r_{\rm cut})$}

Since $X_{ST}$ is positive for both nucleons and $\Delta$'s, 
the total $B_{q q \pi}^{\rm tot}$ will also be positive for
values of $r_{\rm cut}$ greater than 0.7 fm.
That is, pion exchange between quarks produces a repulsive force within these baryons.

To evaluate this contribution to the baryon energy, we must fix the value of $g_{qq\pi}$, as 
noted after Eq.(\ref{eq:PPS}).  To do this, we refer to results from the 
{\it extended} QDCSM \cite{xtnded}
and some general features relating the chiral coupling of pions to the axial current, $f$, to the 
coupling, $g$, for the nominally equivalent (at low energies) pseudoscalar form (used here for 
its relative calculational simplicity).  The usual relation at the nucleon level is 
\Be
   g_{NN\pi} = f_{NN\pi} \frac{2 M_{N}}{m_{\pi}}     \label{eq:ccreln}
\Ee
where the value of $f_{NN \pi}^{2}/(4\pi) = 0.075$, as extracted from fits to nucleon-nucleon 
scattering \cite{friar,nij}.  The factor of $2 M_{N}$ compensates for the fact that in the 
pseudoscalar coupling, $\bar{\psi}\gamma_{5}\psi$, the upper and lower components 
of the Dirac wave function are coupled, with the lower being reduced, in momentum space, 
by a factor of order $m_{\pi}/(E_{N}+M_{N}) \sim m_{\pi}/(2 M_{N})$.  The momentum 
scale is set by $m_{\pi}$ and the kinetic energy of the nucleon is small. 

In the extended QDCSM \cite{xtnded}, (which, as we noted previously, is a 
non-relativistic quark model of nucleon 
structure and nucleon-nucleon interactions extended from the QDCSM by the 
addition of pion exchange between quarks separated by a minimum distance), 
the relation between $f_{NN\pi}$ and $f_{qq\pi}$ is modified from the simple 
group-theoretic factor (5/3) by a small correction due to the quark delocalization 
within the nucleon, which accounts for the nucleon not being a point-like field 
lacking internal structure, viz., 
\Be
   f_{qq\pi}^2 = \frac{9}{25} f_{NN\pi}^2 e^{-m_{\pi}^2 b^2 /2}  \ .   \label{eq:fqqpi}
\Ee
Here $b \approx 0.6$ fm describes the width of a single quark wave function in the nucleon and 
so provides for the experimentally determined root-mean-square 
size of the overall matter distribution in the nucleon.
Since $m_{\pi} b \approx 0.4$, this correction is of order 10\%, so that 
\Be
   f_{qq\pi}^2/4\pi  \approx 0.025  \ .  \label{eq:fqqpinumer}
\Ee

Nonetheless, this value is not appropriate to our relativistic description 
here, as it has been extracted for non-relativistic wave functions. The 
effect is similar to the one evidenced in the right-hand-side of Eq.(\ref{eq:ccreln}), 
as  our lower wave function components are suppressed relative to the 
upper components by a factor of order $m_\pi/(E_q + m_q)$, the 
relevant ratio corresponding to what we saw above for the nucleon. Since the quarks 
here are highly relativistic, but with an energy of order one-third of the 
mass of the nucleon, the suppression is therefore not as great as for the 
case of the very non-relativistic nucleon.  Since this suppression  is already 
present in the scale of our lower wave function components, it follows that 
the appropriate value to use is 
\Be
   g_{qq\pi}^2/4\pi \approx  0.025 \times \left(\frac{E_q + m_q}{m_{\pi}}\right)^2  
   		 \approx   0.17	\ , \label{eq:fqqpifin}
\Ee
where we have taken $E_{q} = 361.6$ MeV and $m_{q} \simeq 0$, as in the LARQ. Note that 
this is approximately two orders-of-magnitude smaller than the 
value for nucleons, $g_{NN\pi}^2/4\pi$, which suggests a more 
valid perturbation theory.

The function $I(r_{\rm cut})$ reaches its minimum near 
$r_{\rm cut} = 1.05$ fm, where its value is $-0.492$ MeV. Multiplying 
together the various factors in Eq.\ (\ref{eq:fullBtot}), the maximum 
contribution for this exchange energy is
\Be
   B_{q q \pi}^{\rm tot} = 5.02 {\rm\ MeV,\  for\  the\  nucleon,} \quad
   B_{q q \pi}^{\rm tot} = 1.00 {\rm\ MeV,\  for\  the\ } \Delta . \label{eq:maxBtot}
\Ee
This contribution to the self-energy produces a small change in mass, compared 
to rest masses of these baryons. It is comparable to electromagnetic corrections, 
which have been ignored at this level. 

This allows us to conclude that the fitted parameters of the LARQ model for the 
nucleon and $\Delta$ baryons are not significantly affected by inclusion of this 
pion-exchange contribution between the quarks. The only remaining concerns 
are related to quark self-energy corrections, which we will deal with in following 
section.

For larger values of $r_{\rm cut}$, such as the 1.4 fm cutoff advocated in 
Ref.\ \cite{nij}, this contribution is even smaller. It is also interesting 
that, for a value of $r_{\rm cut} \approx$ 0.7 fm (as suggested by the MIT bag 
model \cite{MITbag}), this contribution is exceptionally small, since 
$I(r_{\rm cut})$ crosses the $x$-axis near that value.

However, none of this implies that the predictions of the LARQ model for nuclei,
such as the deuteron, will be unchanged by this pion-exchange contribution.
There, significant cancellations affect the net value of the binding energy. Hence 
it is still necessary to extend these calculations to the nuclear cases, and to determine 
the change in this pion-exchange interaction energy as the quark wave function 
changes in each particular nucleus.

%%%%%

\section{Pion-loop contribution to the quark self-energy} \label{sec:piloop}

In addition to the exchange of a pion between different quarks in the hadron, one must
also consider, at the same level, the one-loop contribution to the quark's self-energy.
(See Fig.\ \ref{fig:SEvsExch}.)
We first make an approximation that the major contribution to the loop integral 
for this self-energy comes when the intermediate quark state is given by the bound state 
used in the previous section.
This approximation neglects the contributions of the excited bound-state quark wave functions.
It is similar to one made by Inoue et al. \cite{Inoue}, but we improve
on that work by keeping our calculation fully relativistic.

We then compare the above bound-state approximation with two different approximations, 
where we use a {\it free} massless quark propagator for the intermediate quark state.
One version uses a sharp momentum cutoff, neglecting all pion momenta greater than $k_{\rm cut}$.
The other uses a monopole form factor to cut off, more smoothly, pion momenta greater than its
parameter, $\Lambda$.

As we shall see, all three of these approximations give corrections on the order
of a few MeV, i.e., are also small in comparison with the nucleon and $\Delta$ rest masses.

Comparison of these three approximations weighs heavily on the question of whether 
or not this contribution can be calculated sufficiently reliably, so that the change that 
develops in a nucleus can be reasonably determined. 

\subsection{Bound, Ground-state Approximation}

We begin by evaluating the first approximation to the loop correction.
For a quark labeled ``1'' and a quark in the intermediate state labeled ``2'',
\Be
   B_{qq\pi}^{SE} =  \frac{g^2_{q q \pi}}{4\pi} 
	< \mbox{\boldmath$\tau$} \cdot \mbox{\boldmath$\tau$} > 
	I^{SE}(r_{\rm cut}) \  , \label{eq:BSE}
\Ee
where  $< \mbox{\boldmath$\tau$} \cdot \mbox{\boldmath$\tau$} > = 3$ and
\Be
   I^{SE}(r_{\rm cut}) = \sum_{s_2} \int d^3 r_1 \int d^3 r_2 \;
     (\bar{\psi}_1 i \gamma _5 \psi_2) \; \frac{e^{-m_\pi r}}{r} \Theta(r - r_{\rm cut})\;
       (\bar{\psi}_2 i \gamma _5 \psi_1) \ . \label{eq:ISE}
\Ee
Again, the $\Theta$-function assures that $r = |{\bf r}_1 - {\bf r}_2| \geq r_{\rm cut}$ .
For the $S$-wave quark wave functions of Eq.\ (\ref{eq:psicolvec}) we find
\Be
   (\bar{\psi}_1 i \gamma _5 \psi_2) =
        \left[ \; \psi_a^*(r_1) \psi_b(r_2) \; \chi_{s_1}^\dagger 
	      \mbox{\boldmath$\sigma$} \cdot {\bf \hat{r}}_2 \chi_{s_2} +
	 \psi_b^*(r_1) \psi_a(r_2) \; \chi_{s_1}^\dagger 
	      \mbox{\boldmath$\sigma$} \cdot {\bf \hat{r}}_1 \chi_{s_2} \; \right] 
      \label{eq:psi12loop}
\Ee
and 
\Be
   (\bar{\psi}_2 i \gamma _5 \psi_1) =
        \left[ \; \psi_a^*(r_2) \psi_b(r_1) \; \chi_{s_2}^\dagger 
	      \mbox{\boldmath$\sigma$} \cdot {\bf \hat{r}}_1 \chi_{s_1} +
	 \psi_b^*(r_2) \psi_a(r_1) \; \chi_{s_2}^\dagger 
	      \mbox{\boldmath$\sigma$} \cdot {\bf \hat{r}}_2 \chi_{s_1} \; \right] 
      \label{eq:psi21loop} \ .
\Ee
Multiplying Eqs.\ (\ref{eq:psi12loop}) and (\ref{eq:psi21loop}) and
using completeness of the spin function, $\sum_{s_2} \chi_{s_2} \chi_{s_2}^\dagger = 1$,
we find
\Bea
     \sum_{s_2} (\bar{\psi}_1 i \gamma _5 \psi_2)  (\bar{\psi}_2 i \gamma _5 \psi_1) 
     &=& 2 \;\left[ \psi_a(r_1) \psi_b(r_1) \; {\bf \hat{r}}_1 \cdot {\bf \hat{r}}_2 \;
		\psi_a(r_2) \psi_b(r_2)  \right.\nonumber \\
     && \quad\quad\quad \left. + |\psi_a(r_1)|^2 |\psi_b(r_2)|^2 \right] \ , 
     \label{eq:fourpsis}
\Eea
where we have made use of the reality of the wave functions, 
the anticommutator $\{\sigma_\alpha,\sigma_\beta\} = \delta_{\alpha \beta}$, 
and the interchange symmetry/equivalence of $r_1$ and $r_2$ to combine terms.

With the pion Green's function factor, $e^{-m_\pi r}/r$, the integral over the first
term in Eq.\ (\ref{eq:fourpsis}) is two times $ I(r_{\rm cut})$, 
which we defined in the previous section and calculated to have the form 
given in Eqs.\ (\ref{eq:IexchTrace}) and (\ref{eq:Isum}).  
Thus, the self energy for a single quark from a pion loop consists of two terms,
\Be
   B_{qq\pi}^{SE} = 6 \; (\frac{g^2_{q q \pi}}{4\pi}) \;
      [ \;  I(r_{\rm cut}) + \widetilde{I}(r_{\rm cut}) \;] \ . 
      \label{eq:BSE3terms}
\Ee
The {\it new} contribution  to $B_{qq\pi}^{SE}$, $\widetilde{I}(r_{\rm cut})$, 
comes from the last term in Eq.\ (\ref{eq:fourpsis}).
For a three-quark baryon, the total self-energy correction from pion loops will be
three times $B_{qq\pi}^{SE}$.

We can evaluate $\widetilde{I}$ in a way very similar to that for the more complicated $I$.
\Bea
   \widetilde{I}(r_{\rm cut}) &=& 
       \int d^3 r_1 \int d^3 r_2 \;  \psi_b^2(r_1) \; \psi_a^2(r_2) \; 
	\frac{e^{-m_\pi r}}{r} \Theta(r - r_{\rm cut}) \nonumber \\
      &=& 4\pi N^4 \kappa \int_0^\infty x_1^2 dx_1 \; \phi_b^2(x_1) \; 
	\widetilde{J}(x_1, x_{\rm cut}) 
        \label{eq:Ibbaa} \ ,
\Eea
where
\Bea
   \widetilde{J}(x_1, x_{\rm cut}) &=& \int d^3 x_2 \; 
	\frac{e^{-\mu x}}{x} \Theta(x - x_{\rm cut})\; \phi_a^2(x_2) \nonumber \\
      &=& \int_{x_{\rm cut}}^\infty x^2 dx \; \int d\Omega_x \; 
          \frac{e^{-\mu x}}{x} \; \Theta(x-x_{\rm cut}) \;
	  \phi_a^2(|{\bf x}_1 - {\bf x}|) \ . 
	  \label{eq:Jaadef}						
\Eea
Note that $\widetilde{I}$ is positive and there will be some cancellation in 
Eq.\ (\ref{eq:BSE3terms}) with $I$, which is negative for values of $r_{\rm cut}$ of interest.
How much cancellation depends on the relative sizes of $I$ and $\widetilde{I}$.

Proceeding as in the previous section, with the Gaussian expansion of $\phi_a(x)$ defined
in Eqs.\ (\ref{eq:normpsihat}-\ref{eq:normzn}), we find, after some algebra,
\Bea
   \widetilde{J}(x_1,x_{\rm cut}) &=& \frac{1}{2 x_1} \sum_{k_1 k_2} a_{k_1} a_{k_2} 
         \left( \frac{\pi}{c_{k_1 k_2}} \right)^{3/2} 
	 e^{\mu^2/(4 c_{k_1k_2})} \times \nonumber \\
         & & \quad\quad\quad\left\{ e^{-\mu x_1} \erfc[t_{k_1 k_2}^{(+)}(x_{\rm cut})] -  
	         e^{+\mu x_1} \erfc[t_{k_1 k_2}^{(-)}(x_{\rm cut})] \right\}
	  \ . \label{eq:Jaadone}
\Eea
The quantities $c_{k_1 k_2}$ and $t_{k_1 k_2}^{(\pm)}(x_{\rm cut})$ are the same as above, defined before
Eq.\ (\ref{eq:psiuplosum}) and in Eq.\ (\ref{eq:tkkpm}).
Inserting $\widetilde{J}$ in Eq.\ (\ref{eq:Ibbaa}), we have finally
\Be
   \widetilde{I}(r_{\rm cut}) = 2\pi N^4 \kappa
      \sum_{k_1 k_2} a_{k_1} a_{k_2}
           \left( \frac{\pi}{c_{k_1 k_2}} \right)^{3/2} e^{\mu^2/(4 c_{k_1 k_2})} 
	\widetilde{I}_{k_1 k_2}(x_{\rm cut}) \ , \label{eq:finalIbbaa}
\Ee
where
\Be
   \widetilde{I}_{k_1 k_2}(x_{\rm cut}) = \int_0^\infty x_1 dx_1 \; {\phi_b}^2(x_1)
        \left\{ e^{-\mu x_1} \erfc[t_{k_1 k_2}^{(+)}(x_{\rm cut})] -  
	         e^{+\mu x_1} \erfc[t_{k_1 k_2}^{(-)}(x_{\rm cut})] \right\} \ . \label{eq:Ikk}
\Ee

We have again evaluated this sum of one-dimensional integrals using Mathematica.
It is a monotonically falling function ranging from 11.88 MeV at $r_{\rm cut} = 0.6$ fm
to 0.275 MeV at $r_{\rm cut} = 1.8$ fm.
For $r_{\rm cut} = 1.0$ fm, $\widetilde{I}$ = 6.17 MeV, quite a bit 
larger than $I$ at this value of $r_{\rm cut}$ (-0.487 MeV).
Thus the sum of the integrals in the square bracket in 
Eq.\ (\ref{eq:BSE3terms}) is dominated by the $\widetilde{I}$ term.
The solid curve in Fig.\ \ref{fig:gsBSE} shows the dependence of $B_{qq\pi}^{SE}$ 
on $r_{\rm cut}$.
(The dashed curve shows the dominant contribution due to $\widetilde{I}(r_{\rm cut})$.)

For $r_{\rm cut} = 1.4$ fm that square bracket has a value of
1.436 MeV and the {\it total} value of the quark self-energy loops is
$B_{qq\pi}^{{\rm tot,} \; SE}$ is 4.39 MeV.
This is the contribution of all {\it three} quarks in a nucleon (or $\Delta$)
for this value of $r_{\rm cut}$.
In comparison, for the nucleon, the exchange energy from pion exchange {\it between} 
the three quarks at the same value of $r_{\rm cut}$ is, from Eq.\ (\ref{eq:fullBtot}), 2.58~MeV.
If this value were not much changed by the inclusion of higher energy intermediate states, 
it would be clear that the LARQ parameters are not significantly affected by the addition of 
``long-distance'' pion exchanges. We next test this conclusion by using a free propagator 
approximation for the intermediate state.

% \mbox{\boldmath$\tau_1$}

%\vspace{1 in}

\subsection{Free Propagator Approximation}

To compare with the above approximation we now evaluate, in four-momentum space, the pion-loop
self-energy correction using a free-quark propagator in the intermediate state.
This energy is,
for a bound quark with four-momentum $p_\mu = (E,|{\bf p}|)$, $E = 0.362$ GeV,
an  intermediate pion with four-momentum $k_\mu$,
and a massless intermediate free quark with four-momentum $(p - k)_\nu$,
\Bea
   i B^{\rm SE}_{\rm free} &=& 3 g_{qq\pi}^2
     	\sum_s \int \frac{d^4 p}{(2\pi)^4} \int \frac{d^4 k}{(2\pi)^4}  \; \bar{\psi}_s(p) i \gamma_5 \;
     	\frac{\gamma^\nu (p-k)_\nu}{[(p-k)_\mu (p-k)^\mu+ i\epsilon]} \times \nonumber \\ 
     	& & \quad\quad\quad\quad\quad\quad\quad\quad\quad\quad\quad
     	\quad\quad\quad\quad\quad \frac{1}{(k_\mu k^\mu - m_\pi^2 + i\epsilon)} \;
      	i \gamma_5 \psi_s(p)   \nonumber  \\
      	&=& 3 g_{qq\pi}^2  \;
	\sum_s \int \frac{d^4 p}{(2\pi)^4} \ \bar{\psi}_s(p) \gamma_5 \gamma^\nu \gamma_5 {\psi}_s(p) \;
         	 J_\nu(p_\mu p^\mu) \label{eq:Bfree}
\Eea
The factor of 3 comes from the isospin factor,
 $<\mbox{\boldmath$\tau$} \cdot \mbox{\boldmath$\tau$}>$.

The loop integral $J_\nu$ defined in Eq.\ (\ref{eq:Bfree}) is 
\Be
   J_\nu(p_\mu p^\mu) = \int \frac{d^4 k}{(2\pi)^4} \; 
	  \frac{(p-k)_\nu}{[(p-k)_\mu (p-k)^\mu+ i\epsilon]} \;
      \frac{1}{(k_\mu k^\mu - m_\pi^2+ i\epsilon)}  \equiv J(p_\mu p^\mu) \; p_\nu \ , 
	 \label{eq:defJnu} 
\Ee
since $p_\nu$ is the only available four-vector for defining the $\nu$-component.
Its coefficient is 
\Be
   J(p_\mu p^\mu) = \frac{1}{p_\nu p^\nu}\int \frac{d^4 k}{(2\pi)^4} \ 
	\frac{p_\mu p^\mu - p_\mu k^\mu}{[p_\mu p^\mu - 2 p_\mu k^\mu + k_\mu k^\mu + i \epsilon]} \ 
    \frac{1}{(k_\mu k^\mu - m_\pi^2+ i \epsilon)} \ ,  \label{eq:defJ}
\Ee
which will be evaluated below.

With $\bar{\psi}_s(p) = \psi_s^\dagger(p) \gamma_0$  and the factor of $p_\nu$ from $J_\nu$,
the Dirac matrices in Eq.\ (\ref{eq:Bfree}) reduce to
\Be
   \gamma_0 \gamma_5 \ \gamma^\nu p_\nu \ \gamma_5 = 
      \left( \begin{array}{cc}
         	  -p_0 		&  \mbox{\boldmath$\sigma$} \cdot {\bf p} \\
	   \mbox{\boldmath$\sigma$} \cdot {\bf p} & -p_0
	   \end{array} \right)  \ , \label{eq:gams}
\Ee
which is to be inserted between the spinors  $\psi_s^\dagger(p)$ and $\psi_s(p)$.
We will see presently that this allows the spin sum over $s$ to become trivial.

So, there are two tasks: to determine the momentum-space wave functions $\psi_s(p)$
and to evaluate $J(p_\mu p^\mu)$.

\subsubsection{The Momentum-Space Bound-State Wave Functions}

We need  the four-momentum-space $\psi_s(p)$ corresponding to the bound-state Dirac wave function 
$\psi({\bf x})$ given by Eqs.\ (\ref{eq:psicolvec}), (\ref{eq:psinorm}), (\ref{eq:psihata}),
and (\ref{eq:psihatb}).
To do this, we must add a time-dependence to $\psi({\bf x})$, i.e., we need the four-space Fourier 
transform of $\psi({\bf x}) e^{-iEt}$.

The Fourier transform of the time dependence, $e^{-iEt}$, is a bit tricky,
as one needs to avoid the integration over the square of a $\delta$-function,
$\delta(p_0 - E)$, in the integrals for $B^{\rm SE}_{\rm free}$.  
However, normalizing instead in a box of time of size $T_E$, such that
$n_E = E T_E/(2\pi)$ is an integer, we find the normalized Fourier transform 
in time to be
\Be
   \psi(p_{0n}) = (2 \pi T_E)^{1/2} \; \delta_{n,n_E} \ . \label{eq:normzdPsip0}
\Ee
Further, because of the discretization, integrals over $d p_0/(2\pi)$  are replaced by sums
involving the discrete eigenvalues $p_{0n} = 2 \pi n/T_E$,
\Be
   \int_{-\infty}^\infty \frac{d p_0}{2\pi} \; \psi^\dagger(p_0) f(p_0) \psi(p_0) \rightarrow 
      \frac{1}{T_E} \sum_{n = -\infty}^\infty \psi^\dagger(p_{0n}) f(2\pi n/T_E) \; \psi(p_{0n}) 
      = f(E)
      \ , \label{eq:inttosum}
\Ee

The Fourier transform of the spatial wave function $\psi_s({\bf x})$ is more straightforward.
Dropping the isospinor (which is already incorporated in the $<\mbox{\boldmath$\tau$}\cdot\mbox{\boldmath$\tau$}>$
factor in Eq.\ (\ref{eq:Bfree}) above), we have from Eq.\ (\ref{eq:psicolvec})
\Be
   \psi({\bf p}) = \int d^3 r \; e^{-i{\bf p}\cdot{\bf r}} \psi({\bf r})
      = N \left( \frac{2\pi}{\kappa} \right)^{3/2} \left( \begin{array}{c}
           \tilde{\phi}_a(q) \chi_s  \\
	   - \mbox{\boldmath$\sigma$} \cdot {\bf \hat{p}} \; \tilde{\phi}_b(q) \chi_s
	   \end{array} \right) \  , \label{eq:FTpsiofp}
\Ee
where $q = |{\bf p}|/\kappa$ is dimensionless and
\Bea
   \tilde{\phi}_a(q) &=&  \sum_k  \frac{a_k}{\tilde\mu_k^{3/2}} 
                      e^{-q^2/2 \tilde\mu_k} \ , \nonumber\\
   \tilde{\phi}_b(q) &=& q \sum_k \frac{b_k}{\tilde\mu_k^{5/2}} 
                      e^{-q^2/2 \tilde\mu_k} \  \label{eq:hatpsip}
\Eea
are also dimensionless real functions.

With Eq.\ (\ref{eq:FTpsiofp}),  the wave-function dependence in Eq.\ (\ref{eq:Bfree})
becomes, using Eq.\ (\ref{eq:gams}),
\Be
   \sum_s \psi_s^\dagger(p) \gamma_0 \gamma_5 \gamma^\nu p_\nu \gamma_5 \psi_s(p)  = 
	- N^2 \left( \frac{2\pi}{\kappa} \right)^{3} \left[ p_0 \;  \tilde{\phi}_a^2(q)
	+ 2 |{\bf p}| \; \tilde{\phi}_a(q)\tilde{\phi}_b(q)  
	+ p_0 \; \tilde{\phi}_b^2(q) \right] \ , \label{eq:psistructure}
\Ee
the spin-sum having become trivial, as advertised above.
(The Pauli spin matrices again disappear from the problem since 
$\{\sigma_i {\hat{p}}_i, \sigma_j {\hat{p}_j}\}
	= 2 \delta_{ij} \; {\hat{p}}_i {\hat{p}_j} = 2$.)
The three terms in Eq.\ (\ref{eq:psistructure}) are in decreasing order of importance.

\subsubsection{The Loop Integral \label{subsub:loop}}

The loop integral $J(p_\mu p^\mu)$ is not the usual self-energy one-loop correction 
for a free particle for two reasons. First, 
the external quark is a bound-state particle, requiring an integration over its
four-momentum, $p_\mu$.
The integration over $dp_0$ is trivial, using Eq.\ (\ref{eq:inttosum}), 
so the work to be done is in evaluating the triple integral over $d |{\bf p}|$.
(The four-momentum $p_\mu$ is time-like for $|{\bf p}| < E$ and space-like otherwise.)
Secondly, the integration over the pion four-momentum $k_\mu$ is ``cut,'' corresponding to
the above coordinate-space integrations being evaluated for 
$r = |{\bf r}_1 - {\bf r}_2| \geq r_{\rm cut}$.  
Hence we want to restrict the integration over the pion three-momentum, $|{\bf k}|$, to be 
from 0 to $k_{\rm cut} = \hbar c/r_{\rm cut} \approx 0.2$ GeV/c.

We first carry out the integral over $d k_0$ (from $-\infty$ to $+\infty$) 
using the Feynman trick, 
\Be
   \frac{1}{ab} = \int_0^1 dx \; \frac{1}{[\;xa + (1-x)\;b\;]^2} \ . \label{eq:Ftrick}
\Ee
Letting $a = (p-k)_\mu (p-k)^\mu + i\epsilon$, $b = k_\mu k^\mu - m_\pi^2 + i\epsilon$, 
and four-vector $l_\nu = k_\nu - x p_\nu$, we find
\Be
    J(p_\mu p^\mu) =  \int_0^1 dx (1 - x) \int_{\rm cut} \frac{d^4 l}{(2\pi)^4} \; 
	\frac{1}{[l_\mu l^\mu - R^2(p,x)]^2 + i\epsilon} \ . \label{eq:Jdblpole}
\Ee
Here we have dropped terms linear in $l_\mu$ (they vanish) and have defined 
\Be
   R^2(p,x) = - x(1-x) p_\mu p^\mu + (1-x) m_\pi^2 \  \label{eq:Rdef} \  , 
\Ee
Note that $R^2$, despite its appearance, is not positive definite; as a quadratic in $x$ 
it can go negative for $m_\pi^2/ x  + {\bf p}^{2}  < E^{2}$.  
As we will see, this, among other things, leads to some imaginary parts in the 
evaluation of $J(p_\mu p^\mu)$.

Continuing, we have
\Be
   J(p_\mu p^\mu) = \int_0^1 dx \; (1 - x) \int_{\rm cut} \frac{d^3 l}{(2\pi)^3}
		\int_{-\infty}^{+\infty} \frac{d l_0}{(2\pi)} \; \frac{1}{(l_0^2 - S^2 + i\epsilon)^2} \ .
		\label{eq:dl0intgl}
\Ee
where
\Be
   S^2(p,x,k) = |\mbox{\boldmath $l$}|^2 + R^2(p,x) \ . \label{eq:Sdef}
\Ee
Using Cauchy's integral theorem for integrating over the double pole at 
$l_0 = -S + i\epsilon$, we find
\Be
   J(p_\mu p^\mu) = \frac{i}{8} \int_0^1 dx \; (1 - x) \int_{\rm cut} \frac{d^3 l}{(2\pi)^3} \;
	\frac{1}{[|\mbox{\boldmath $l$}|^2 + R^2(p,x) ]^{3/2}} \  . \label{eq:Jdxd3l}
\Ee

Unfortunately, the limits for the integration over $d^3 l$ are angle dependent and
complicated.    Thus we go back to an integration over $d^3 k$.
In the following we abbreviate the three-vector magnitudes $|{\bf p}|$ and
$|{\bf k}|$ by $p$ and $k$, respectively.
\Be
      J(E^2 - p^2) = \frac{i}{32 \pi^2} \int_0^1 dx \; (1 - x) \; K(p,x) \ , \label{eq:JdxoverK}
\Ee
where
\Bea
   K(p,x) &=& \int_0^{k_{\rm cut}} k^2 dk \; L(p,x,k)  \  , \label{eq:KdkoverL} \\
   L(p,x,k) &=& \int_{-1}^1 d \cos\theta \; \frac{1}{(A - B\cos\theta)^{3/2}}  \ ,
	\label{eq:Ldef} \\
   A(p,x,k) &=&  k^2 + x^2 p^2 + R^2(p,x)  \label{eq:Adef} \ ,\\
   B(p,x,k) &=& 2 x p k   \geq 0  \label{eq:Bdef} \ .
\Eea
Note that the functions $A$ and $B$ here are different from the $A_{k_1,k_2}$ and
$B_{k_1,k_2}$ defined in Sec.\ IIC.
More importantly, $A(p,x,k)$ can be less than $B(p,x,k)$ and even can become negative
for certain small values of $p$ and $k$ when $0 < x < 1$.
Thus the integrals $K(p,x)$ and  $L(p,x,k)$ can be complex in some regions of 
their variables.
These imaginary parts arise in situations when the initial bound quark,
sufficiently off-shell, can decay into an on-shell pion (with three-momentum ${\bf k}$)
and a free (massless) quark.
Such decays will not contribute to the self-energy being calculated, since the quark is 
confined to the nucleon in reality.
Therefore we discard any imaginary parts of $K(p,x)$ and $L(p,x,k)$ below.

Doing the integration over $\cos(\theta)$, we obtain
\Be
   L(p,x,k) = 
	\frac{2}{B} \left[ \frac{1}{\sqrt{A-B}} - \frac{1}{\sqrt{A+B}} \right]			
	 \ , \label{eq:Lpxk}
\Ee
where, as just discussed, it is understood that we only use the real part 
of $L$ in the later integrations.
In the limit that $B \rightarrow 0$, $L \rightarrow 2/A^{3/2}$, so there is
no real singularity at $k = 0$.

The integral over $dk$ defining $K(p,x)$ can also be done analytically, yielding
\Bea
   K(p,x) &=& D(p,x) + E(p,x) \ , \label{eq:KforAgtBgt0} \\
   D(p,x) &=& \frac{\sqrt{(k_{\rm cut} - p x)^2 + R^2(p,x)} -
		\sqrt{(k_{\rm cut} + p x)^2 + R^2(p,x)}}{2px} \ , \label{eq:Ddef} \\
   E(p,x) &=& \frac{1}{2} \; \ln\left[ \frac{F_+(p,x) F_-(p,x)}{R^2(p,x)} \right] \ , 	
	\label{eq:Edef} \\
   F_\pm(p,x) &=&  k_{\rm cut} \pm p x + \sqrt{(k_{\rm cut} \pm p x)^2 + R^2(p,x)}
	\ . \label{eq:Fpmdef}
\Eea
In the limit that $x \rightarrow 0$, 
$D(p,x) \rightarrow -2 k_{\rm cut}/\sqrt{k_{\rm cut}^2 + m_\pi^2}$, and there 
is no singularity at $x = 0$. Also, noting that $R^2(p,1) = 0$, one might be 
concerned about the logarithmic divergence at $x = 1$ of $K(p,x)$.
However, this singularity is moderated by the factor of $(1 -x )$ in the 
integral over $dx$ in Eq.\ (\ref{eq:JdxoverK}).  For the subsequent integration 
over $p$, the $1/p$ in $D(p,x)$ will be cancelled by the $p^2$ in the $d^3 p$ 
volume element.

Again, we only take the real part of $K(p,x)$ in the later integrations over $x$ and $p$.
For certain values of $0 < x < 1$ and (small) $k_{\rm cut}$, the structure of $K(p,x)$
can be {\it quite} complex. The argument of the logarithm in $E(p,x)$ can pass through 
zero and even become complex in several places.  Not only does the denominator 
$R(p,x)$ become negative, but so can the arguments of the square roots in $F_\pm$.
An example for $k_{\rm cut} = 0.100$ GeV/c showing the real and imaginary parts 
of $K$ is given in Fig.\ \ref{fig:comK2}.

It is also possible to do integrations over $dx$ in Eq.\ (\ref{eq:JdxoverK}) analytically,
but the resulting expressions are extremely cumbersome and not very informative.
Thus we have elected to do the remaining integration for $J(E^2-p^2)$ over $dx$ 
numerically,  along with the necessarily numerical integration over $dp$ to obtain 
the value of $B^{\rm SE}_{\rm free}$.

\subsubsection{Combining These Results}

 From Eqs.\ (\ref{eq:Bfree}), (\ref{eq:inttosum}), (\ref{eq:psistructure}), 
(\ref{eq:JdxoverK}), and (\ref{eq:KdkoverL}-\ref{eq:Fpmdef}), we have, for a single quark,
\Bea
   B^{\rm SE}_{\rm free}(k_{\rm cut}) &=& - \frac{3}{2} \frac{g_{qq\pi}^2}{4\pi} \; 
	\frac{N^2}{\kappa^3} \; \int_0^\infty p^2 dp
	\left\{ E \left[\tilde{\phi}_a^2(q) + \tilde{\phi}_b^2(q) \right] + 
		2 p \; \tilde{\phi}_a(q)\tilde{\phi}_b(q) \right\}  \nonumber \\
	& & \qquad\qquad\qquad \times \int_0^1 dx \; (1-x) K(p,x) \ . \label{eq:Bintdpdx}
\Eea
As just stated, we have done the integrations over $dp$ and $dx$ numerically.
We find that, as expected, there is a monotonic decrease in 
$B^{\rm SE}_{\rm free}(k_{\rm cut})$ as the 
cutoff parameter $k_{\rm cut}$ gets smaller (i.e., as $r_{\rm cut}$ gets larger), 
as shown in Fig.\ \ref{fig:Bfreekcut}.
To make the comparison with Fig.\ \ref{fig:gsBSE} easier, we have plotted it
as a function of $\hbar c/k_{\rm cut}$ in fm.
The fact that $B$ becomes slightly negative about 1.4 fm is probably an artifact of
our sharp cutoff approximation; see the next subsection for another version
of the free propagator approximation in which the self-energy stays positive.

If $r_{\rm cut}$ is identified with  $\hbar c/k_{\rm cut}$ (i.e., with
proportionality factor 1.0), then the self-energy in this case is {\it very}
small for values of $r_{\rm cut}$ greater than 1.0 fm.
The proportionality factor, however, need not be 1.0.

For the nucleon (or the $\Delta$), the result of Eq.\ (\ref{eq:Bintdpdx}) 
is to be multiplied by three.

\subsubsection{Alternative Loop Integral with a Monopole Form Factor Cutoff}

The loop integral $J_\nu$ in Eq.\ (\ref{eq:defJnu}) can also be ``cut''  by introducing
a monopole cutoff factor $-\Lambda^2/(k_\mu^2 - \Lambda^2 + i\epsilon)$ in its integrand.
This factor provides convergence to an otherwise logarithmcally divergent 
integral.
It is normalized so its value is 1 when $k_\mu^{\ 2}$ goes to 0.
The momentum parameter $\Lambda$ is (roughly) linearly related to the sharp 
cutoff $k_{\rm cut}$, and its proportionality factor will be discussed later.

With the monopole factor in the loop integral above, $J_\nu(p_\mu^2)$, we rename it as
\Bea
   I_\nu(p_\mu^2,\Lambda) &=& \int \frac{d^4 k}{(2\pi)^4} \; 
	\frac{(p-k)_\nu}{[(p-k)_\mu^{\ 2} + i\epsilon\,]} \ 
	\frac{1}{(k_\mu^{\ 2} - m_\pi^2+ i\epsilon)} \;
	\frac{-\Lambda^2}{(k_\mu^{\ 2} - \Lambda^2 + i\epsilon)} \nonumber \\
	&=& \frac{-\Lambda^2}{\Lambda^2 - m_\pi^2}
	\int \frac{d^4 k}{(2\pi)^4} \; \frac{(p-k)_\nu}{[(p-k)_\mu^{\ 2} + i\epsilon\,]} \ 
	\left[ \frac{1}{(k_\mu^{\ 2} - m_\pi^2 + i\epsilon)} -
		\frac{1}{(k_\mu^{\ 2} - \Lambda^2 + i\epsilon)} \right]
		 \nonumber \\
	&\equiv& \frac{-\Lambda^2}{\Lambda^2 - m_\pi^2} \;
	\left[ J_\nu(p_\mu^{\ 2}, m_\pi^2) - J_\nu(p_\mu^{\ 2}, \Lambda^2) \right] \ . 
	\label{eq:defInu} 
\Eea
Note that $J_\nu(p_\mu^{\ 2}, m_\pi^2)$ is just what we called $J_\nu(p_\mu^{\ 2})$
earlier.

As before, with $M^2 = m_\pi^2$ and $\Lambda^2$,
\Be
   J_\nu(p_\mu^{\ 2}, M^2) = J(p_\mu^{\ 2}, M^2) \; p_\nu \ . \label{eq:JMnuJpnu}
\Ee
The evaluation of the scalars $J(p_\mu^{\ 2},M^2)$ proceeds much as before, 
but in fact the final integrations become simpler.  
With the same Feynman trick,
\Bea
    J(p_\mu^{\ 2},M^2) &=&  \int_0^1 dx (1 - x) \int \frac{d^4 l}{(2\pi)^4} \; 
	\frac{1}{[l_\mu l^\mu - R^2(p,x,M)]^2 + i\epsilon} \ , \label{eq:JMdblpole} \\
  	R^2(p,x,M) &=& - x(1-x) p_\mu^{\ 2} + (1-x) M^2 \  \label{eq:RMdef} \  .
\Eea
Integrating over the double pole in $k_0$, we now find
\Be
   J(p_\mu^{\ 2},M^2) 
   	= \frac{i}{8} \int_0^1 dx (1 - x) \int \frac{d^3 l}{(2\pi)^3} \;
	\frac{1}{[ l^2 + R^2(p,x,M) ]^{3/2}} \  , \label{eq:JMdxd3l}
\Ee
where $l = |\mbox{\boldmath $l$}|$.
Again, $ l^2 + R(p,x,M)$ can become negative for certain values
of $p$ and $x$.
We continue ignoring any imaginary parts of $J(p_\mu^{\ 2},M^2)$ when we 
calculate $I_\nu(p_\mu^{\ 2})$.

Doing the trivial angular integrations and integrating with respect to $dl$, 
we find the simple result that
\Bea
   I_\nu(p_\mu^{\ 2},\Lambda) &=& \frac{i p_\nu}{16 \pi^2} 
	\left( \frac{\Lambda^2}{\Lambda^2 - m_\pi^2} \right) 
	\int_0^1 dx \; (1 - x) \;\ln \left[ \frac{\Lambda^2 - x p_\mu^{\ 2}}
	{m_\pi^2 - x p_\mu^{\ 2}} \right] \nonumber \\
	&=& \frac{i p_\nu}{16 \pi^2} \; K(p,\Lambda) \  . \label{eq:InuIntdx}
\Eea
where 
\Be
   K(p,\Lambda) =  \frac{\Lambda^2}{\Lambda^2 - m_\pi^2} \;
	\int_0^1 dx \; (1 - x) \;\ln \left[ \frac{a - x}{b - x} \right] 
		\ , \label{eq:Klambdadef}
\Ee
and we have defined
\Bea
   a(p,\Lambda) &=&  \Lambda/p_\mu^{\ 2} = \Lambda/(E^2 - p^2) \ , \label{eq:apdef} \\
   b(p,m_\pi)   &=&  m_\pi^2/p_\mu^{\ 2} = m_\pi^2/(E^2 - p^2) \ . \label{eq:bpdef} \ 
\Eea
[The function $K(p,\Lambda)$ here is different from that defined by Eqs.\ (\ref{eq:JdxoverK}) 
and (\ref{eq:KforAgtBgt0}).]
The logarithm in $K$ has two singularities, when either $\Lambda^2 - x p_\mu^{\ 2}$ or
$m_\pi^2 - x p_\mu^{\ 2}$ equals zero.
However, logarithmic singularities are integrable, so that is not a real problem.
Assuming $\Lambda > m_\pi$, the argument of the logarithm in the integrand
first goes negative when \mbox{$p^2 <E^2-m_\pi^2$}.
As already discussed above, the resulting imaginary contributions to $K(p,\Lambda)$ and
hence $I_\nu(p_\mu^{\ 2})$ are irrelevant for our self-energy calculation, and 
they will be discarded.

Carrying out the integration over $x$ in Eq.\ (\ref{eq:Klambdadef}), we find
\Bea
   K(p,\Lambda) &=& \frac{\Lambda^2}{\Lambda^2 - m_\pi^2} \;
	\left\{ a - b + \;\ln\left( \frac{a-1}{b-1} \right) 
	+ \left( a^2-2 a \right) \,\ln\left( \frac{a-1}{a} \right) \right. \nonumber \\
	& & \left.  \qquad\qquad\qquad
	-\;\left(b^2-2b \right) \,\ln\left( \frac{b-1}{b} \right) 
	\right\} \ , \label{eq:Kresult}
\Eea
bearing in mind that $a$ and $b$ are functions of $p$.
This result is valid even when the arguments of the logarithms go negative.
For example, Fig.\ \ref{fig:Kofp} shows the real and imaginary parts of 
$K(p,\Lambda)$ as a function of $p$ for a value of $\Lambda = 0.350$ GeV/c.
Note the smoother behavior of the real part as $p$ goes through the singularity
at $p =\sqrt{E^2 - m_\pi^2}$, where $I_\nu$'s  value is $\ln(\Lambda/m_\pi)$.
For smaller values of $\Lambda$ the real part is more sharply peaked at 
$\sqrt{E^2 - m_\pi^2}$ and the width of the positive imaginary ``spike" 
between the two singularities is narrower.

 From Eqs.\ (\ref{eq:Bfree}), (\ref{eq:inttosum}), (\ref{eq:psistructure}), and 
(\ref{eq:InuIntdx}), we have, for a single quark,
\Be
   B^{\rm SE}_{\rm free}(\Lambda) = - \frac{3}{2} \frac{g_{qq\pi}^2}{4\pi} \;  \int_0^\infty p^2 dp
	\left\{ E \left[\tilde{\phi}_a^2(q) + \tilde{\phi}_b^2(q) \right] + 
		2 p \; \tilde{\phi}_a(q)\tilde{\phi}_b(q) \right\} K(p,\Lambda) \ . \label{eq:Bintdp}
\Ee
We have carried out  the integration over $dp$ numerically.
There is, in this approximation, again a monotonically decreasing dependence on the 
inverse of the cutoff parameter $\Lambda$, as shown in Fig.\ \ref{fig:BfreeLambda}.
(Here we have chosen $x$-axis variable to be $\hbar c/\Lambda$, in units of fm,
to compare better with the ground state approximation shown in Fig.\ \ref{fig:gsBSE}.)

The comparison of Fig.\ \ref{fig:BfreeLambda} with Fig.\ \ref{fig:Bfreekcut}
was at first surprising to us, but the results for the value of $B^{SE}_{\rm free}$
are comparable if $\Lambda \approx 2.5 k_{\rm cut}$, as shown in Fig.\ \ref{fig:shiftedLambda}.

\subsection{Comment on renormalization}

The reader familiar with field theory may be concerned that we have not carried 
out mass and wave function renormalization subtractions of the quantity calculated 
above. We have not done so because the renormalized field theoretic degrees of 
freedom in the LARQ are massless, current quarks. To the extent that pion self 
energy corrections must be applied, we view these as implementing the construction 
of constituent quarks, which have revised properties, as remarked in the Introduction. 

Nevertheless, the fundamental approach of the LARQ was to carry out calculations 
with current quarks and with minimal modeling of QCD. From that point of view, the 
pion loop self-energy correction should be subtracted and the quark wavefunction 
correspondingly renormalized to recover the current quarks of the model. If this is 
carried out then, since the calculations above produce relatively small values, this 
makes clear that the changes after renormalization must be very small. We may 
then further infer that the changes induced by the nuclear distortions of the nucleon 
quark wavefunctions produce differences between a nucleus and $A$ nucleons that 
are even smaller. 

It thus appears to be reasonable to conclude rather strongly that, as for the pion 
exchanges between quarks analyzed in Sec.\ref{sec:piexch}, these effects on the 
parameters of the LARQ due to including pion-quark coupling are negligible. The 
only question remaining for the LARQ is how much the pion interactions between 
quarks from different nucleons affects the total binding energy of the nucleus. We 
will address this question for $^{3}$He and $^{4}$He in a later publication. 

%\pagebreak
%\vspace{0.5 in}

\section{Relation to other calculations and conclusions}  

The calculations in the previous two sections correspond to calculation of the 
two graphs in Fig.\ \ref{fig:SEvsExch}. 
The exact result for Fig.\ \ref{fig:SEvsExch}b requires a summation over all 
intermediate excited states of the quark within the confining potential or, 
equivalently, use of the in-medium quark propagator rather than the free 
propagator. 
However, as the two results of using only the lowest bound state and the 
free propagator are both small, 
we conclude that we have a reasonable estimate of 
the long-distance pionic contributions to the mass of the nucleon and of the 
Delta within the LARQ quark model of these states. 

Estimates of the pionic (pion cloud) contributions to these states have also 
been made using chiral pertubation theory ($\chi$PT).  These calculations 
are restricted (so far) to the contributions shown in Fig.\ \ref{fig:SEvsExch} 
in a form which ignores the substructure and uses the low energy constants (LEC) 
that are extracted from data using $\chi$PT. It should be noted that the LARQ 
calculation also implicitly includes all other intermediate states in the pion loop 
calculations of $\chi$PT -- all $N^{*}$, $\Delta^{*}$, etc. states. From $\chi$PT, 
one expects their contributions to be small and decreasing with increasing mass. 
As such, it is interesting to compare $\chi$PT results with those of the LARQ. 

Unfortunately, there does not seem to be much agreement among the various 
$\chi$PT calculations as to the result. For example, Ref.~\cite{otherChiPT} obtains a 
large correction ($\approx 200$ MeV) for the total nucleon self-energy, using only 
the ground state baryon as the intermediate state, (where the substructure is 
used to compute an effective vertex function for the $\pi NN$ coupling rather than 
working with the substructure directly as done here).  Other papers~\cite{AWT} 
include $\Delta$ and $N^{\star}$ intermediate state resonances with ambiguous 
results. 

In conclusion, we find that the contribution of this additional interaction to the mass 
of the nucleon (and the $\Delta$-baryon) is small. The scale of effects that we have 
found is comparable to the expected scale for electromagnetic corrections, which 
were not included in the LARQ and so contribute to its overall scale uncertainty. 
We conclude from this that the LARQ is stable under this change, in the sense that 
significant changes to the model parameters are not required to maintain a match to 
data on state masses. We also infer that inclusion of these pion exchange contributions 
in calculations of nuclear states will not significantly affect the quark structure of the 
nuclei considered, although it could improve the agreement with actual binding 
energies. We will report results of those calculations elsewhere. 

This work has been supported by the Department of Energy under contract W-7405-ENG-36.
We acknowledge useful conversations and correspondence with J.\ L.\ Friar,
J.\ N.\ Ginocchio, A.\ K.\ Kerman, M.\ M.\ Nieto, Jialun Ping, K.\ E.\ Schmidt, 
and A.\ Steiner.

\section{Appendix: Group Theoretic Calculation of the Spin-Isopin Sum}

An elegant (and relatively quick) calculation of the $X_{ST}$ factor uses some group 
theory techniques.
First, rewrite $X_{ST}$ in terms of the spin and isospin angular momentum operators,
\Be
X_{ST} = 8 \sum_{i \neq j} 
     <S,M_S | \; {\bf s}^{(i)} \cdot {\bf s}^{(j)} \; |S,M_S>
     <T,M_T | \; {\bf t}^{(i)} \cdot {\bf t}^{(j)} \; |T,M_T> \  .
     \label{eq:newXST} 
\Ee
The reason for doing this is because the three quarks are in a symmetric representation 
of the group SU(4), for which the generators are $s_\alpha$, $t_\alpha$, and $s_\alpha t_\beta$
in the fundamental representation.
For our case of three symmetric quarks, these generators become 
\Be
   S_\alpha = \sum_{i=1}^3 s_\alpha^{(i)}, \quad T_\alpha = \sum_{i=1}^3 t_\alpha^{(i)}, \quad 
   {\rm and } \  \ G_{\alpha\beta} = \sum_{i=1}^3 s_\alpha^{(i)} t_\beta^{(i)} \ . \label{eq:SU4gens}
\Ee

The sum in $X_{ST}$ can be evaluated from the quadratic Casimir invariant $C_2^{(4)}$ for the 
symmetric SU(4) representation for $N$ and $\Delta$, whose Young 
tableau has ``overhang numbers''  $(\lambda,\mu,\nu) = (3,0,0)$.  
For a general SU(4) representation \cite{C2SU4}
\Bea
   C_2^{(4)} &=& \frac{1}{4} [3 \lambda(\lambda+4) + 4 \mu(\mu+4) + 3 \nu(\nu+4)
                  + 4 \mu(\lambda+\nu) + 2 \lambda\nu]  \label{eq:C2gen} \\
      &=& 63/4 , \quad {\rm for} \  (\lambda,\mu,\nu) = (3,0,0) \ . \label{eq:C2for300}
\Eea
The $C_2^{(4)}$ can also be expressed in terms of the total spin and isospin quantum
number and the scalar formed from the $G_{\alpha\beta}$ as \cite{JNG}
\Be
   C_2^{(4)} = S(S+1) + T(T+1) + 4 <G_{\alpha\beta} G_{\alpha\beta}> \ , \nonumber 
   \label{eq:C2inSTG}
\Ee
Breaking up product in the last term into terms with $i=j$ and those with $i \neq j$,
\Bea
   G_{\alpha\beta} G_{\alpha\beta} 
		  &=& \sum_{i=1}^3 s_{\alpha}^{(i)} t_{\beta}^{(i)} s_{\alpha}^{(i)} t_{\beta}^{(i)} +
		       \sum_{i \neq j=1}^3 s_{\alpha}^{(i)} t_{\beta}^{(i)} s_{\alpha}^{(j)} t_{\beta}^{(j)} 
		       \nonumber \\
		  &=& \sum_{i=1}^3 ({\bf s}^{(i)} \cdot {\bf s}^{(i)}) ({\bf t}^{(i)} \cdot {\bf t}^{(i)}) +
		       \frac{1}{8} X_{ST} \ . \label{eq:splitGG}
\Eea
The sum on $i$ in the first term is easy,
\Be
   \sum_{i=1}^3 \frac{3}{4} \times \frac{3}{4} = \frac{27}{16} \ .
\Ee
Thus
\Bea
   X_{ST} &=& 8 <G_{\alpha\beta} G_{\alpha\beta}> - \frac{27}{2} \nonumber \\
          &=& 2 \left[ <C_2^{(4)}> - S(S+1) - T(T+1) \right] - \frac{27}{2} \nonumber \\
          &=& 18 - 2 \left[ S(S+1) + T(T+1) \right]  \ , \label{eq:finalXST}
\Eea
using the value of 63/4 for $<C_2^{(4)}>$ given in Eq.\ (\ref{eq:C2for300}).
The spin-isospin sums for $N$ and $\Delta$ are therefore
\Be
   X_{1/2,\; 1/2} = 15 \ , \quad\quad\quad\quad\quad  X_{3/2,\; 3/2} = 3 \ ,
   \label{eq:XSTvalues}
\Ee
in agreement with the values given in Sec.\ {\ref{sec:XST}.

%%%%%%%------------------------------ 

\newpage

%Figures

\begin{figure*}[ht] %Fig. 1
\includegraphics{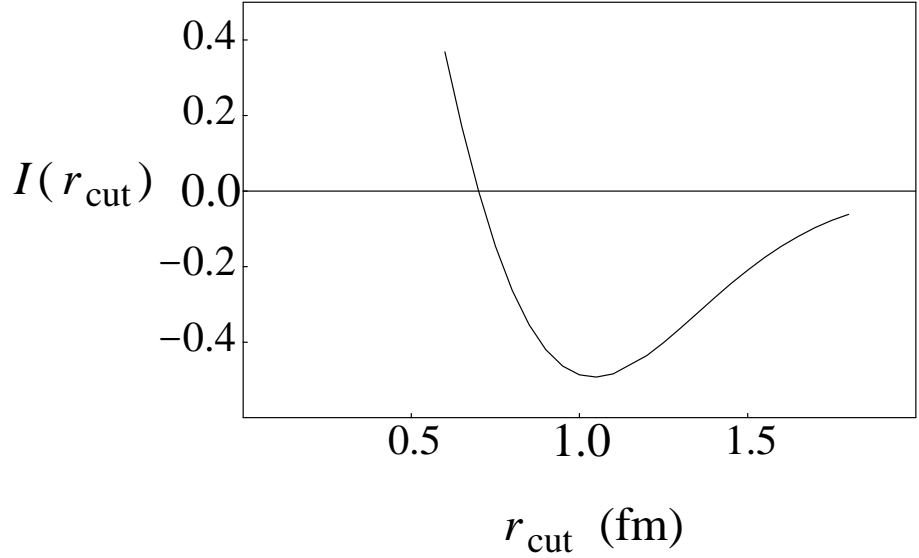}
\caption{Variation of integral for exchange between different quarks, $I$, 
as a function of $r_{\rm cut}$.
The $x$- and $y$-axes have units of fm and MeV, respectively. 
\label{fig:Ircut}}
\end{figure*} 

\begin{figure*}[ht] %Fig. 2
\includegraphics{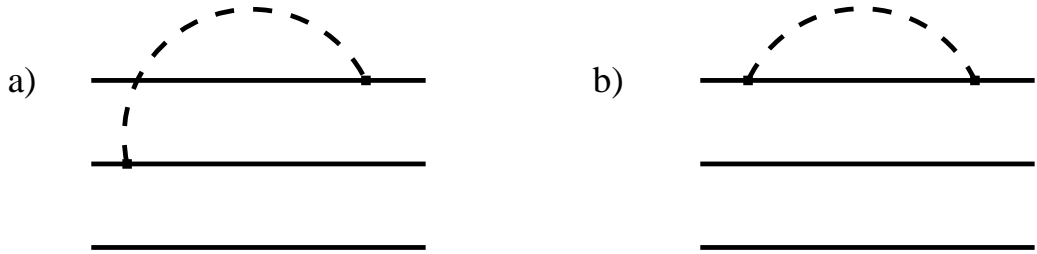}  
%SEvsExch.eps} 
\caption{Both the self-energy pion-loop, and pion exchange between separate 
quarks, appear as self-energy corrections to the nucleon or the $\Delta$ when 
the internal substructure of the baryons is ignored, as in Chiral perturbation Theory.   
\label{fig:SEvsExch}}
\end{figure*} 

\begin{figure*}[ht] %Fig. 3 
\includegraphics[width=.7\textwidth,height=0.5\textwidth]{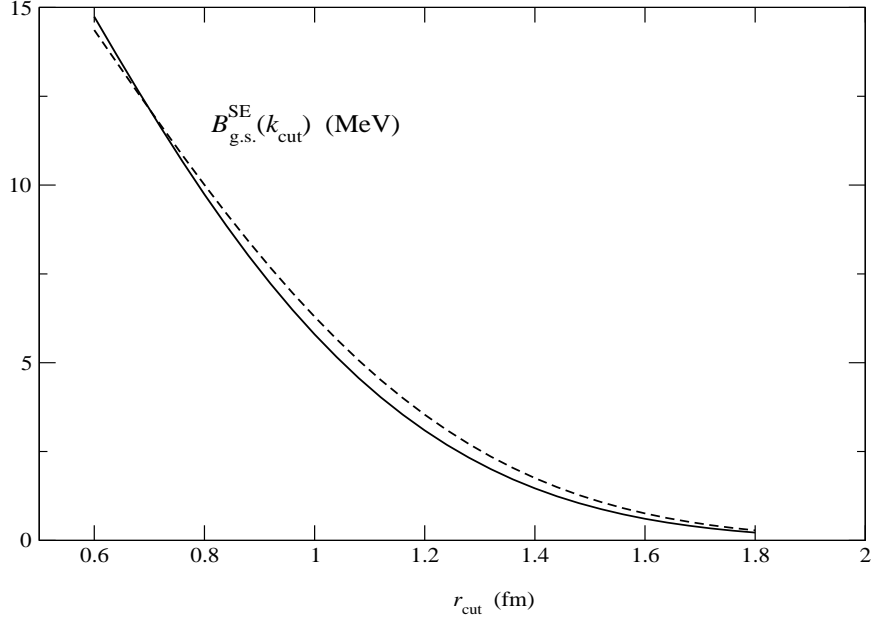}
\caption{The solid curve shows the pion-loop self energy correction for a single quark
as function of $r_{\rm cut}$.  The dashed curve shows the dominant contribution coming from
$\widetilde{I}(r_{\rm cut})$, the difference coming from the smaller $I(r_{\rm cut})$ 
displayed in Fig.\ \ref{fig:Ircut}.
\label{fig:gsBSE}}
\end{figure*} 

\begin{figure*}[ht] %Fig. 4
\includegraphics{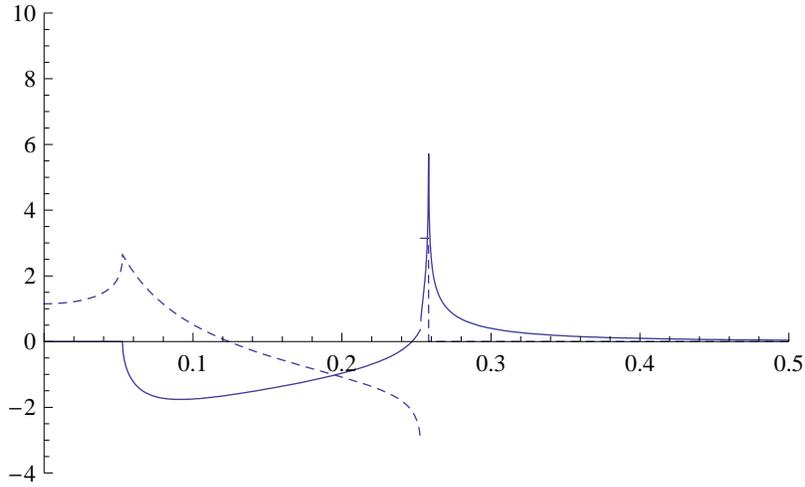} 
\caption{$K(p,x)$ for $x=0.3$ and $k_{\rm cut} = 0.100$ MeV/$c$ as a function of $p$ (in MeV/$c$).  
The solid curve displays the real part of $K$, the dashed curve is for the imaginary part.  
The ``spike'' in the imaginary part near $p = 0.26$ MeV/$c$ is actually a narrow 
flat-topped pedestal with value $+\pi$. 
\label{fig:comK2}}
\end{figure*}

\begin{figure*}[ht] %Fig. 5
\includegraphics[width=.7\textwidth,height=0.5\textwidth]{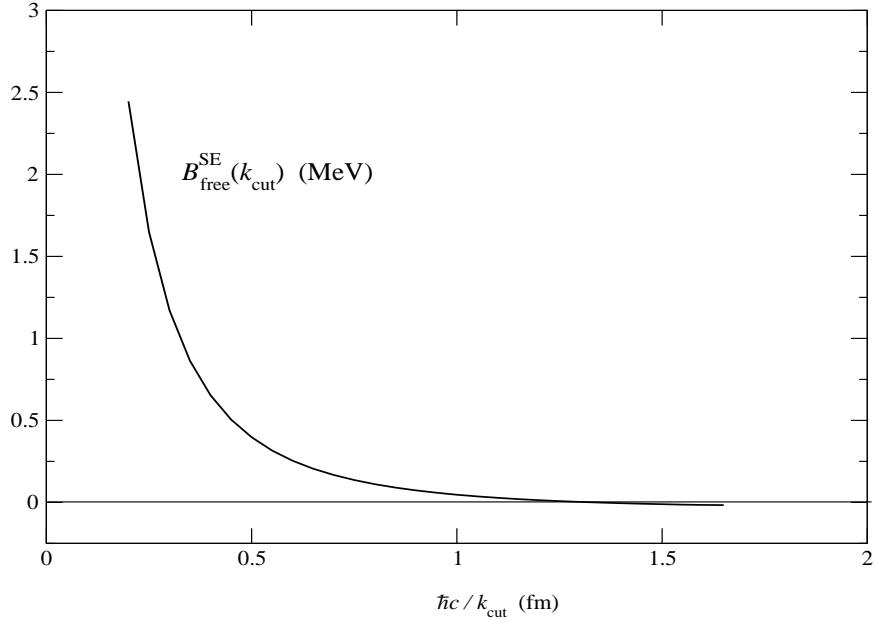}
\caption{The free propagator approximation for the self-energy of a single quark with a
pion-loop correction with sharp cutoff, $k_{\rm cut}$. 
$B^{\rm SE}_{\rm free}(k_{\rm cut})$ is shown as a function 
of $\hbar c/k_{\rm cut}$ for comparison with Fig.\ \ref{fig:gsBSE}.
} \label{fig:Bfreekcut}
\end{figure*}

\begin{figure*}[ht] %Fig. 6
\includegraphics{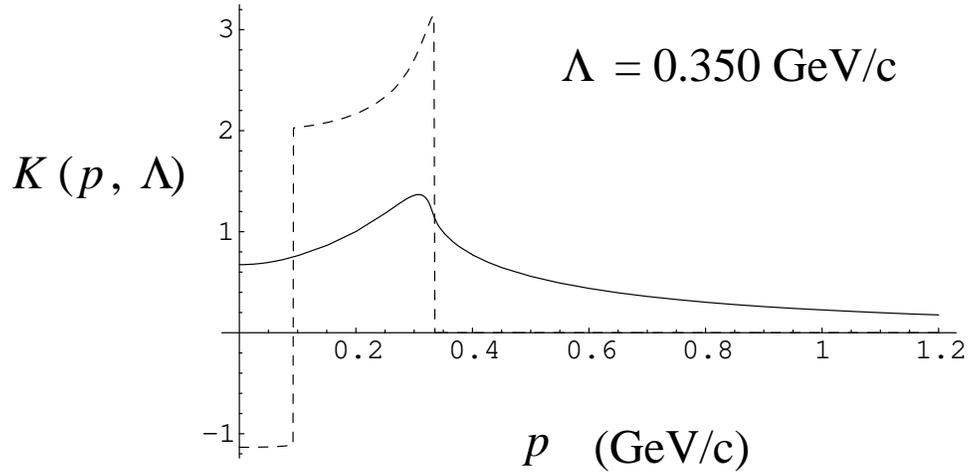} %/Users/silbar/Public/Kofp.eps}
\caption{The real (solid curve) and imaginary(dashed curve) parts of $K(p,\Lambda)$ 
for a value of $\Lambda = 0.350$ GeV/$c$, close to, but less than, the bound quark energy, 
$E = 0.362$ GeV. \label{fig:Kofp}}
\end{figure*}

\begin{figure*}[ht] %Fig. 7
\includegraphics[width=.7\textwidth,height=0.45\textwidth]{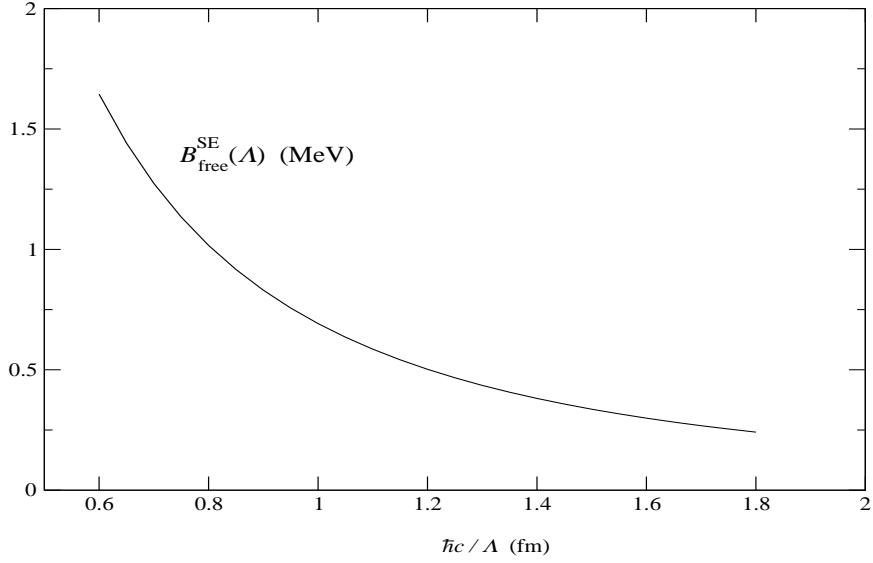} 
\caption{The self-energy contribution for a single quark, $B^{\rm SE}_{\rm free}(\Lambda)$, 
for a pion-quark loop correction using a monopole cutoff parameter $\Lambda$.  
It is shown as a function of $\hbar c / \Lambda$ in fm, for comparison with both
Fig.\ \ref{fig:gsBSE} and Fig.\ \ref{fig:Bfreekcut}.
}  
\label{fig:BfreeLambda} 
\end{figure*}

\begin{figure*}[ht] %Fig. 8
\includegraphics[width=.7\textwidth,height=0.45\textwidth]{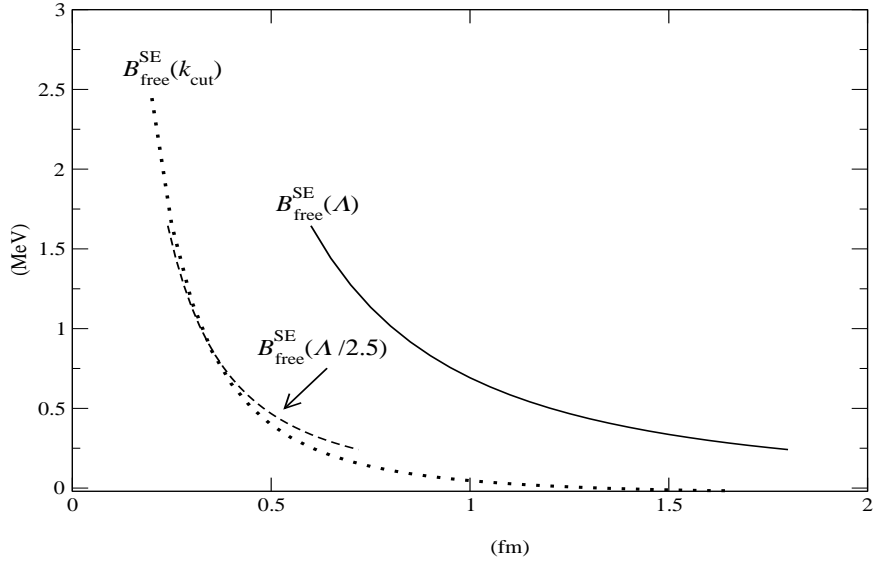} 
\caption{The solid curve shows $B^{\rm SE}_{\rm free}(\Lambda)$, as in Fig.\ \ref{fig:BfreeLambda},
while the dotted curve is $B^{\rm SE}_{\rm free}(k_{\rm cut})$, as in Fig.\ \ref{fig:Bfreekcut}.
The dashed curve indicates how the agreement between these two curves for $\Lambda$ 
reduced by a factor of 2.5. \label{fig:shiftedLambda} }
\end{figure*}

\end{document}